\documentclass[aps,prb,twocolumn,showpacs,amsmath,amssym,
superscriptaddress]{revtex4}
\pdfoutput=1
\usepackage{amssymb}

\usepackage{graphicx}
\usepackage{dcolumn}
\usepackage{color}
\usepackage{verbatim}
\newcommand{\cmo}{CaMn$_7$O$_{12}$}

\begin{document}

\title{First-principles study of structurally modulated multiferroic CaMn$_7$O$_{12}$}

\author{Kun Cao
\footnote{corresponding author, Email address: kun.cao@materials.ox.ac.uk} }
\affiliation{Department of Materials, University of Oxford, Parks Road, Oxford OX1 3PH, United Kingdom}

\author{Roger D. Johnson}
\affiliation{Clarendon Laboratory, Department of Physics, University of Oxford, Parks Road, Oxford OX1 3PU, United Kingdom}

\author{Natasha Perks}
\affiliation{Clarendon Laboratory, Department of Physics, University of Oxford, Parks Road, Oxford OX1 3PU, United Kingdom}

\author{Feliciano Giustino}
\affiliation{Department of Materials, University of Oxford, Parks Road, Oxford OX1 3PH, United Kingdom}

\author{Paolo G. Radaelli}
\affiliation{Clarendon Laboratory, Department of Physics, University of Oxford, Parks Road, Oxford OX1 3PU, United Kingdom}

\date{\today}

\begin{abstract}
We study the electronic and magnetic structures of multiferroic CaMn$_7$O$_{12}$ by first-principle calculations, based on the experimentally determined modulated crystal structure. We confirm the presence of a 3$d$ orbital modulation of the Mn2 (Mn$^{3+}$) sites, previously inferred from the Jahn-Teller crystal distortions. Our results indicate that in the multiferroic phase the magnetic structure of the Mn3 (Mn$^{4+}$) sites is anharmonically modulated \textit{via} orbitally-mediated coupling with the structural modulation, and that the Dzyaloshinskii-Moriya and exchange striction mechanisms contribute equally to the polarization.
\end{abstract}
\maketitle


\section{Introduction}
Research into magnetoferroelectrics, in which improper ferroelectricity is induced by magnetic ordering, has had a significant impact in condensed matter physics and materials science over the last ten years. \cite{cheong07,fiebig05,hur04b,kimura03,goto04,hur04}
These materials have attracted wide attention because of their novel physics and their potential applications in multifunctional devices. One important and currently limiting factor is that, in order for these materials to be useful in technology, the magnitude of the electric polarization must be of the order of that observed in proper ferroelectrics.
The multiferroic CaMn$_7$O$_{12}$ (labelled CMO in the remainder) was shown to exhibit giant magneto-ferroelectricity of magnitude $\sim$ 2.9 mC/m$^{2}$ [\citenum{zhang11,johnson12,perks12}] --- one of the largest measured values of magnetically induced electric polarization.

CMO has a trigonal strucuture (space group $R\overline{3}$) below $\sim$ 400 K, consisting of three symmetry inequiva-
lent manganese sites that we label Mn1, Mn2, and Mn3 --- see Fig.~\ref{fig:CMOMAG}(a).\cite{bochu80} On decreasing the temperature below $T_\mathrm{OO}$=250 K, the average crystal structure adopts an incommensurate modulation (ICM structure) propagating along the hexagonal $c$ axis with structural propagation vector  ${\bf q}_c = (0, 0, 2.077)$ at 150~K. On further cooling, long range magnetic order develops below $T_\mathrm{N1} \sim$90 K with a magnetic propagation vector ${\bf q}_{m1} = (0, 0, 1.037)$(AFM1 phase), followed by a second magnetic phase (AFM2 phase) below $T_\mathrm{N2}$, described by (at least) two propagation vectors ${\bf q}_{m2} = (0, 0, 0.96)$ and  ${\bf q}_{m3} = (0, 0, 1.12)$. Neutron powder diffraction showed that in the AFM1 phase, all spins lie in the $ab$ plane with the two Mn$^{3+}$ sites (Mn1 on the $A'$ perovskite site and Mn2 on the $B$ site) forming ferromagnetic triangular layers that are rotated by 124$^\circ$ with respect to neighbouring layers along the $c$ axis, as described by ${\bf q}_{m1}$. Spins on the Mn$^{4+}$ (Mn3) sublattice are roughly antiparallel to those in neighbouring Mn2 sites, but the precise angle of rotation is poorly constrained by experiments.

Recently, it was proposed that in AFM1, the structural and magnetic modulations lock in a  magneto-orbital helix and that the incommensurate orbital modulation, inferred from the Jahn-Teller distortions, is crucial in stabilizing the chiral magnetic structure.\cite{perks12} In this model, the giant magnetically induced electric polarization would arise primarily from the inverse Dzyaloshinskii-Moriya (DM) effect.\cite{sergienko06}  This picture seems to be contradicted by first-principles calculations, which were performed on an average, \emph{unmodulated} structure of CMO, and led to the proposal that the spin directions of Mn3 are stabilized by strong DM interactions\cite{lu12} rather than orbital modulations. Given this spin structure of Mn3, the electric polarization would then be induced by symmetric exchange striction.\cite{wang07,wang08}
In this paper, we revisit the magnetic structure and magnetoelectric coupling mechanism in CMO using first-principles calculations on a series of
structure models which approximate locally the \emph{modulated}  structure. Our calculations confirm the existence of an orbital modulation as described in our previous paper.\cite{perks12} Furthermore, we propose that the magnetic structure of Mn3 should be modulated anharmonically with two wave vectors ${\bf q}_{m1}$ and $3{\bf q}_{m1}$. Finally, we show that the giant magneto-ferroelectricity observed in experiments is due to equal contributions from both exchange striction and the inverse DM mechanism.

\section{Methods}
Our first-principles calculations are based on the density-functional theory implemented in the Vienna \textit{ab initio} simulations package
(VASP).\cite{kresse93,kresse96} We use the spin-polarized generalized gradient approximation with on-site Coulomb interactions $U$ included for Mn 3$d$
orbitals (GGA+$U$).\cite{liechtenstein95} The projector augmented-wave (PAW)\cite{blochl94} method with a 500 eV
plane-wave cutoff is employed.  A $2 \times 2 \times 4$ Brillouin zone mesh is used. Electric polarization is calculated using the Berry phase method.\cite{king-smith93} 

The magnetic structures of the Mn1 and Mn2 sublattices in the AFM1 phase, which have been well determined by experiment, form the basis for our calculations. The magnetic propagation vector of the AFM1 phase is approximated by the commensurate vector ${\bf q}^\prime_{m1} = (0, 0, 1)$. This approximation results in a 120$^\circ$ rotation (rather than the experimental 124$^\circ$) between neighbouring Mn1 and Mn2 layers (shown in Fig. \ref{fig:CMOMAG}a). Whilst the magnetic structure of Mn1 and Mn2 is fixed, the spin directions of Mn3 ions are allowed to fully relax. The above scheme is employed for all calculations unless otherwise stated.

It is well known that both magnetic and electric properties from GGA+$U$ calculations are very sensitive to the Hubbard parameter $U$. Experiments show that {\cmo} is a bad insulator at room temperature with a band gap $< 0.38$ eV. In our calculations, we fixed $J=1$ eV and tried several different $U$ values. We find that for $U < 3$ eV, the experimental spin configuration becomes metallic, while $U= 3 $ eV produces a small band gap $\sim 0.15 $ eV. There is no other qualitative difference among the results calculated with different U values, so here, unless otherwise stated, we show the results calculated with $U=3$ eV and $J=1$ eV.

\section{structural modulation and orbital helix}
\label{orbital}

\begin{figure}
\centering
\includegraphics[width=3.5in]{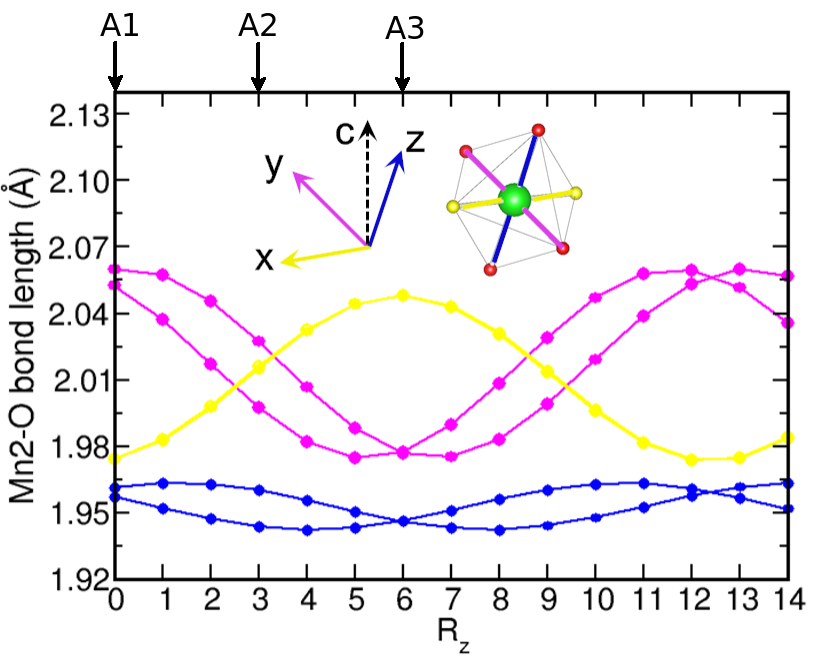}
\caption {Experimental bond-length modulation in CMO : Mn2-O bond lengths along the local octahedral $x$,
$y$ and $z$ (see inset) as a function of $R_z$, plotted across 15 unit cells along the $c$ axis.}
\label{fig:bondmod}
\end{figure}

At 150K, the modulated structure can be characterized by a centrosymmetric four-dimensional space group $R\overline{3}(00 \gamma)$ with propagation vector ${\bf q}_c = (0, 0, 2.077)$.\cite{perks12}  This structural modulation is dominated by variations of Mn2-O bonds, with smaller
variations of Mn1-O and Mn3-O bonds. The MnO$_6$ octahedral crystal field splits Mn $3d$ energy levels into higher {\it $e_g$} and lower {\it $t_{2g}$} states. Since Mn2 has a $3d^4$ configuration (Mn$^{3+}$), the {\it $t_{2g}$} states are fully occupied while the {\it $e_g$} states are partially occupied with parallel spins. The structural modulation leads to local deformations of the oxygen octahedra, which can be characterized by two distortion modes: tetragonal ($Q_3$) and orthorhombic ($Q_2$). $Q_3$  and  $Q_2$ can be further expressed as $Q_3 =(2/\sqrt{6}) (2Z-X-Y)$, $Q_2 =(2/\sqrt{2}) (X-Y)$, where X, Y, and Z correspond to  Mn-O bond lengths in the {\it x},{\it y}, and {\it z} directions defined in the local frame\cite{perks12} (see Fig. \ref{fig:bondmod}). The general distortion can then be written as $|Q \rangle = \cos \alpha |Q_3 \rangle + \sin \alpha |Q_2 \rangle$, where $\alpha$ is a mixing angle. According to the Jahn-Teller effect, the wave function of an $e_g$ state is given by $| \Phi \rangle = \cos (\alpha/2) |3z^2-r^2 \rangle + \sin (\alpha/2) |x^2-y^2 \rangle $, where $\tan(\alpha)=Q_2/Q_3=\frac{\sqrt{3}(X-Y)}{(2Z-X-Y)}$.\cite{goodenough_book} It is easy to show that $\alpha=2 \pi/3$ and $4 \pi/3$ correspond to the occupation of $3x^2-r^2$ and $3y^2-r^2$ orbitals respectively. 

%


%
\begin{figure}
\centering
\includegraphics[width=3.5in]{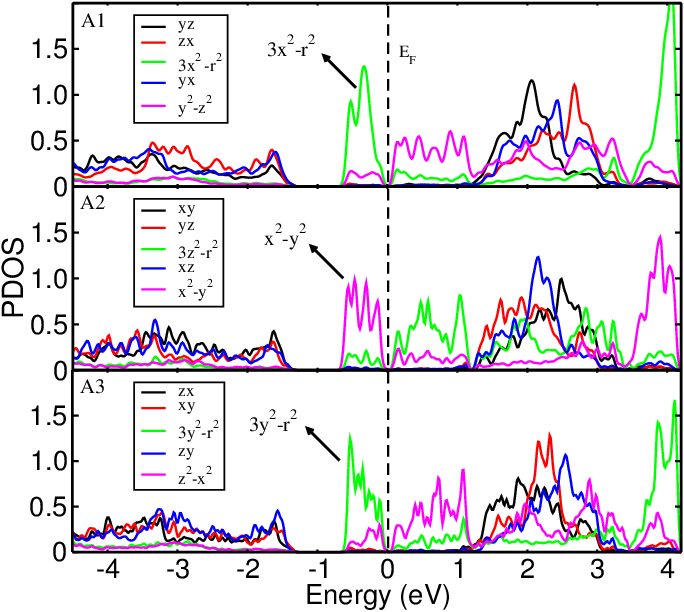}
\caption {Partial density of states for the Mn2 sites of structures A1, A2, and A3. The Fermi energy is at $E_F=0$.}
\label{fig:PDOS}
\end{figure}

In the average (unmodulated) structure, the Mn$^{3+}$O$_6$ octahedra of the Mn2 site are compressed, giving two short bonds along {\it z} and four long bonds in the {\it xy} plane. At variance with this situation, the modulated structure leads to an incommensurate antiphase variation in the bond lengths in the {\it xy} plane, consistent with a rotation of $\alpha$ and therefore a periodic modulation of the occupation of the $3x^2-r^2$ and $3y^2-r^2$ orbitals. The modulation of the Mn2-O bond lengths is shown in Fig. \ref{fig:bondmod}.
Since the propagation vector is incommensurate, it is not possible to study the full structural modulation directly from first-principles calculations, since the corresponding super cell is infinitely long and any realistic commensurate approximation would be computationally prohibitive.
As an alternative, we study a series of models designed to capture the \emph{local} electronic structure in the presence of the modulation.  In this approach, an individual unit cell along the {\it c} direction is selected and then built up as a crystal using periodic boundary conditions. The experimental structural parameters are used, unless otherwise stated. Orbital order is then investigated by determining the orbital occupation of the Mn2 atom centred at (1/2,1/2,1/2), so as to minimise the influence of the artificial unit cell boundary. In order to keep the computational load tractable, we choose to perform calculations on three typical unit cells marked with $R_{z}=0, 3, 6$ in Fig.~\ref{fig:bondmod}, denoted here as A1, A2 and A3, respectively. These points correspond to the two extrema of the modulation (A1 and A3) and the intermediate nodal point (A2). The $\alpha$ values of $A_1$, $A_2$, and $A_3$ structures are 128$^\circ$, 183$^\circ$ and 223$^\circ$, which correspond to $e_g$ occupation with dominant $3x^2-r^2$, $x^2-y^2$ and $3y^2-r^2$ characters respectively. A standard way to understand the local orbital occupation is through calculation of the partial density of states (PDOS). 
The calculated PDOS are shown in Fig.~\ref{fig:PDOS}. From Fig.~\ref{fig:PDOS}, we can see that the $e_g$ orbitals are separated at the Fermi level by a small gap.  The respective $e_g$ orbitals of A1, A2, and A3 clearly have a dominant $3x^2-r^2$, $x^2-y^2$ and $3y^2-r^2$ occupation below the energy gap, respectively, which is in excellent agreement with predictions based on the Jahn-Teller mechanism.

\section{Magnetic structure}
\label{mag}

\begin{figure}
\centering
\includegraphics[width=3.3in]{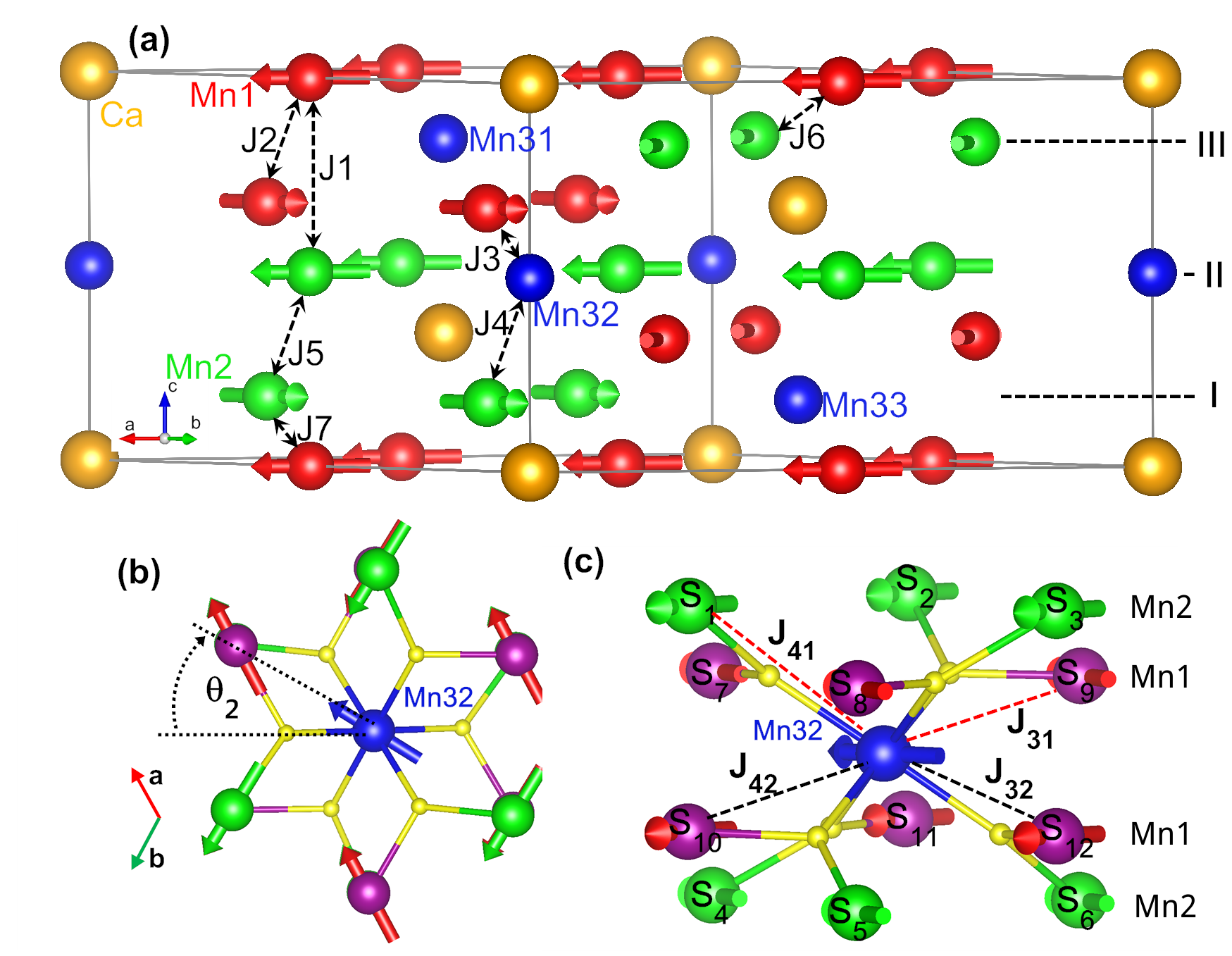}
\caption {(a) Side view of one unit cell of average crystal structure and magnetic configuration. Yellow (large) and blue (small) spheres represent Ca and Mn3 ions respectively. Red (dark) and green (light) spheres with arrows represent Mn1 and Mn2 ions respectively. 
Arrows denote magnetic moments of Mn ions. Nearest neighbour exchange interations are indicated by dotted lines. (b) Top view and (c) side view of one cluster centered at Mn32 atoms. Dotted lines indicate the splitting of $J_3$ and $J_4$ in the modulated structure.}
\label{fig:CMOMAG}
\end{figure}
In addition to the pure electronic structure, magnetic ordering in CMO is also expected to be strongly affected by the structural modulation. To understand this aspect, we performed additional calculations on the series of structure models described above, investigating specifically the coupling between structure and magnetism. In each unit cell of {\cmo}, there are a total of three Mn3 atoms located in different layers, which we denote here as Mn31, Mn32, and Mn33, respectively, as shown in Fig. \ref{fig:CMOMAG}(a). Each Mn3 ion forms a cluster with 6 nearest neighbour (NN) Mn1 and 6 NN Mn2, as illustrated in Fig. \ref{fig:CMOMAG}(b) and (c). The directions of the magnetic moments of Mn31, Mn32, Mn33 with respect to the neighbouring Mn1 and Mn2 layers are expressed through the angles $\theta_1, \theta_2,$ and $\theta_3$. 

We employed a Heisenberg-like model including both NN symmetric exchange (SE) interactions and antisymmetric Dzyaloshinskii-Moriya (DM) exchange interactions. \cite{dzyaloshinsky58,moriya60, lu12} The magnetic interaction associated with one pair of Mn ions joined by a common oxygen ligand  ion can be written as 
\begin{equation}
H= J_{ij} {\bf S}_i \cdot {\bf S}_j+   {\bf D}_{ij} \cdot ({\bf S}_i \times {\bf S}_j),
\label{eq:exchange}
\end{equation}
where the first term is the Heisenberg exchange energy and the second term is the antisymmetric DM exchange energy which originates from spin orbit coupling (SOC). ${\bf S}_i$ and ${\bf S}_j$ are spins of Mn$_i$, Mn$_j$, ${\bf D}_{ij}=\beta{{\bf u}_i \times {\bf u}_j}$ is the DM vector, where ${\bf u}_i$ and ${\bf u}_j$ are vectors connecting the two Mn sites with the oxygen ligand, and $\beta$ is a coupling constant. For NN SE interactions, we use the same notations as in Ref. ~\onlinecite{lu12}. As shown in Fig. \ref{fig:CMOMAG}, there are 7 inequivalent exchange interactions in the average structure $J_1-J_7$, among which $J_3$ and $J_4$ correspond to Mn3-Mn1 and Mn3-Mn2 super-exchange interactions, respectively. In the modulated crystal structure, the exchange interactions are modulated along the $c$ direction. The resulting splitting of $J_3$ and $J_4$ is illustrated for the Mn32 cluster in Fig. \ref{fig:CMOMAG}(c) where in one cluster $J_3$ and $J_4$ are relabelled $J_{31}$, $J_{41}$, and $J_{32}$, $J_{42}$, corresponding to interactions with the upper and lower Mn1 and Mn2 layers, respectively. The DM vectors $\bf{D}$ are denoted using the same subscripts as the SE interactions. According to Eq. (\ref{eq:exchange}), the exchange energy of the Mn32 cluster with the empirical spin directions fixed for Mn1 and Mn2 can be written as

\begin{eqnarray} 
E  & = & J_a \cos\left(\frac{\pi}{3}-\theta_2\right)+J_b\cos\left(\frac{\pi}{3}+\theta_2\right) \nonumber \\
  &   & +D_a\sin\left(\frac{\pi}{3}-\theta_2\right)-D_b\sin\left(\frac{\pi}{3}+\theta_2\right)  \nonumber \\
  & = & \left[\frac{1}{2}(J_a+J_b)+\frac{\sqrt{3}}{2}(D_a-D_b)\right]\cos(\theta_2)  \nonumber \\
  &   & +\left[\frac{\sqrt{3}}{2}(J_a-J_b)-\frac{1}{2}(D_a+D_b)\right]\sin(\theta_2).
\label{eq:energy}
\end{eqnarray}
where $J_a=3(J_{31}+J_{42})$, $J_b=3(J_{32}+J_{41})$, $D_a=3(D^{z}_{31}+D^{z}_{42})$, and $D_b=3(D^{z}_{32}+D^{z}_{41})$. By minimizing the energy with respect to $\theta_2$, we obtain
\begin{equation}
\tan(\theta_2)  =  \frac{\sqrt{3}(J_a-J_b)-(D_a+D_b)}{(J_a+J_b)+\sqrt{3}(D_a-D_b)}.
\label{eq:theta}
\end{equation}
These equations can be generalized to other Mn3 clusters in the whole lattice. 
In the average structure $\theta_1 = \theta_2 = \theta_3= \theta$ is enforced by symmetry, and Mn3 atoms occupy sites with $\bar{3}$ point symmetry, which implies that $J_a=J_b$ and $D_a=D_b$. In the absence of SOC, $D_a=D_b=0$ and Eq. (\ref{eq:theta}) becomes $\tan(\theta_2) = \frac{\sqrt{3}(J_a-J_b)}{(J_a+J_b)}$, from which we can easily obtain $\theta=0$. On the other hand, including SOC will cause a deviation from $0^\circ$ due to DM interactions. Our {\it ab initio} calculations of the average structure give $\theta=0$ without SOC and $\theta=-4.1^\circ$ when SOC is taken into account. This supports the 
validity of the above analysis.

\begin{figure}
\centering
\includegraphics[width=3.3in]{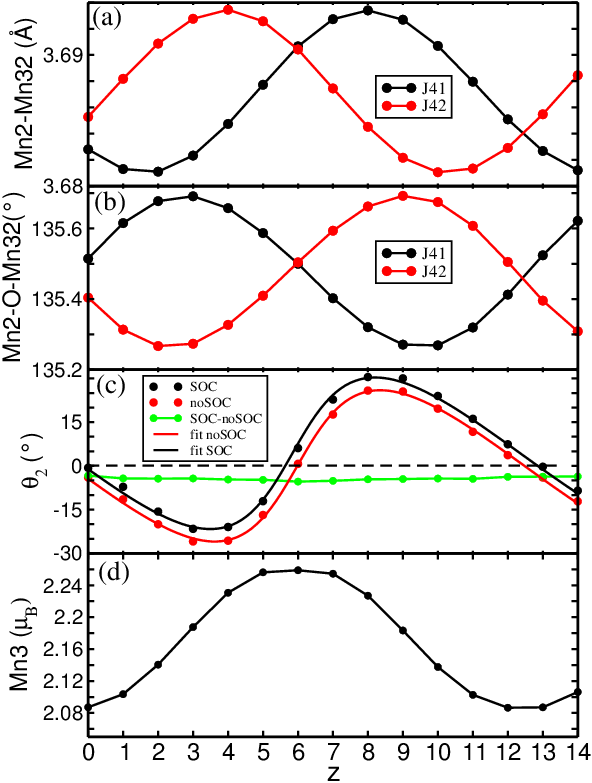}
\caption {(a)(b) Experimental data of modulated Mn32-Mn2 bond lengths and bond angles along the $c$ direction.~\cite{perks12} (c) $\theta_2$ from {\it ab initio} calculations. Red and black lines are fitted to Eq.~(\ref{eq:thetar}). Dashed line is drawn at $\theta_2=0$. (d) The corresponding magnetic moments of the Mn3 ions.}
\label{fig:bond}
\end{figure}

As already stated, in the fully modulated atomic structure, the Mn3-Mn1 and Mn3-Mn2 bonds are modulated along the $c$ direction. As in Sec. \ref{orbital}, we study the modulation of the Mn3 spin direction by performing bulk calculations on periodic structures constructed from unit cells extracted from the modulated structure. Once again, to reduce the impact from artificial neighbouring cells along $c$ direction, we only investigate the Mn32 atom located in the middle layer of each unit cell with $z=1/2$. To illustrate the structural modulation around the Mn32 site quantitatively, Mn32-Mn2 bond lengths and Mn32-O-Mn2 bond angles are plotted in Fig.  
 \ref{fig:bond}(a)(b). It can be seen that both bond lengths and bond angles vary with the wave vector ${\bf q}_c$, as expected. 
The modulation of Mn32-Mn1 has a very similar behavior. We can therefore naturally assume that both $J_a$ and $J_b$ --- see Eq. (\ref{eq:energy}) --- will be modulated with the wave vector ${\bf q}_c$, $i.e.$ $J_a({\bf r})=J_0+\Delta J \sin( {\bf q}_c \cdot  {\bf r}+\phi+\psi)$, $J_b({\bf r})=J_0+\Delta J \sin( {\bf q}_c \cdot  {\bf r}+\phi)$, where $\Delta J $ is the magnitude of the change in $J$ as a result of the modulated structure, $\phi$ is a phase shift and $\psi$ is the phase difference between these two modulations. 
We find that, in the modulated structure, the vectors {\bf D}/$\beta$  connecting the Mn3 layers to the NN Mn2 and Mn1 layers are only slightly modulated and always have the same sign, indicating a weak modulation of both $D_a$ and  $D_b$ for all three Mn3 sites.
The {\it ab-initio} calculated $\theta_2$ are shown in Fig. \ref{fig:bond}(c). We can see that the $\theta_2$ modulation is essentially identical in the presence and in the absence of SOC, indicating that the dominant factor at play is the effect of bond lengths and angles on the Heisenberg exchange. To compare with the average structure, we can consider for example the $A_3$ structure ($z= 6$).  For $A_3$,  $J_a=J_b$, as in the average structure, and hence $\theta_2 \approx 0$ without SOC, as before. The only effect of the SOC is to introduce a constant $\theta_2$  offset of approximately $-4^\circ$,  consistent with the optimized $\theta=-4.1^\circ$ in the average structure.  In order for the offset to be constant along $c$, Eq. (\ref{eq:theta}) imposes that $D_a$ and $D_b$ should also be weakly modulated along the structural modulation.  Consequently, we can safely set $D_a=D_b=D$ and re-write Eq. (\ref{eq:theta}) as
\begin{equation}
\tan\theta_2({\bf r})  =  \frac{\sqrt{3}\left[J_a({\bf r})-J_b({\bf r}) \right]-2D}{J_a({\bf r})+J_b({\bf r})}
\label{eq:thetar}
\end{equation}
By fitting Eq. (\ref{eq:thetar}) to the \textit{ab initio} values of $\theta_2$ in Fig.~\ref{fig:bond}(c), we obtain $\Delta J/J_0 = 0.49$, $\phi=0.40 \pi$, $\psi=0.35 \pi$, $\Delta D/J_0 = 0.08$. The quality of this fit is excellent as can be seen in Fig. \ref{fig:bond}(c). The modulation of $J$ is about half of $J_0$,  preserving the sign of the exchange interactions throughout the whole crystal. To double check the validity of the above analysis, we calculated the magnetic moment magnitudes of Mn32 along the modulation, which are shown in Fig. \ref{fig:bond}(d). It can be seen that the magnetic moments are around 2.18 $\mu_B$ with a slight modulation of amplitude  $\sim0.1 \mu_B$, which agrees very well with experiments.\cite{johnson12}

One important consequence of the $\theta_2$ modulation is to introduce an additional magnetic propagation vector for the Mn3 sublattice. A Fourier analysis of the Mn3 spin structure obtained from Eq.~(\ref{eq:thetar}) gives three modulation vectors ${\bf q}_{m1}$, ${\bf q}_c+{\bf q}_{m1}$ and ${\bf q}_c-{\bf q}_{m1}$. Since ${\bf q}_c=2{\bf q}_{m1}$,  ${\bf q}_c-{\bf q}_{m1}={\bf q}_{m1}$ so that the first and third propagation vectors are equal, whereas ${\bf q}_c+{\bf q}_{m1}=3{\bf q}_{m1}$. This leads us to predict that the magnetic structure of the Mn3 sublattice in the AFM1 phase is \emph{anharmonically modulated} with two propagation vectors, ${\bf q}_{m1}$ and $3{\bf q}_{m1}$,due to orbitally-mediated coupling with the structural modulation. In this context, the term ``anharmonic'' refers purely to the presence of higher-order harmonics of the magnetic propagation
vector {\bf q}$_{m1}$. As we have seen, this conclusion is independent on whether or not the SOC is taken into account, since this does not change the periodicity of the $\theta_2$ modulation.

\section{Electric polarization}
\label{polar}

\begin{figure}
\centering
\includegraphics[width=3.3in]{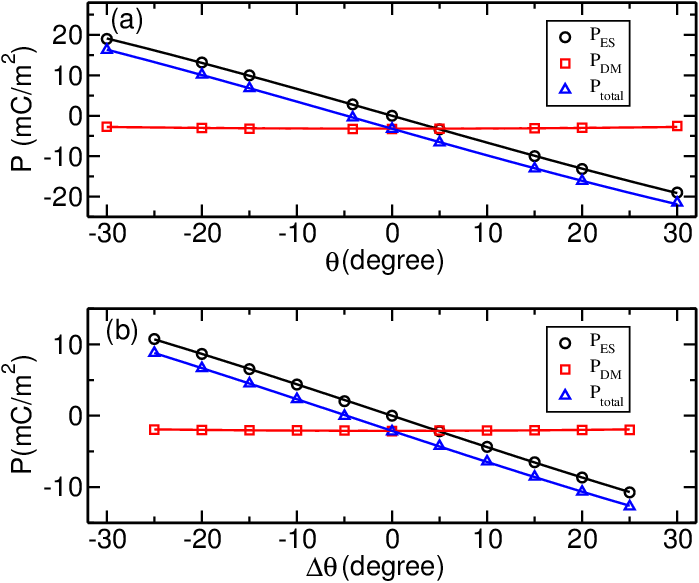}
\caption {Electric polarization of (a) the average structure and (b) the CM structure. Symbols denote results obtained from {\it ab initio} calculations. Lines represent fits to Eq.~(\ref{eq:PES}) and Eq.~(\ref{eq:PDM}).}
\label{fig:polar}
\end{figure}

In this section, we describe our calculations of the electric polarization in the AFM1 phase of \cmo\, performed with the same parameter setup as in Section \ref{mag}. 
We initially performed calculations on the \emph{average} atomic structure for comparison, where the magnetic structure is only characterized by the spin direction $\theta$ of Mn3, as before. The electric polarization calculated as a function of the angle $\theta$ is shown in Fig. \ref{fig:polar}(a). The distinct contributions to the polarization from the symmetric exchange striction ($P_{\rm{ES}} $) and inverse DM effects ($P_{\rm{DM}} $) are identified by turning on and off the SOC. Fig. \ref{fig:polar}(a) shows that $P_{\rm{ES}} \propto -\theta$, while $P_{\rm{DM}} $ is approximately constant.  As already remarked,\cite{perks12}  $P_{ES}$ must be zero for $\theta=0$ because magnetic structures with opposite chiralities are related by a global rotation in spin space.  These calculated results can be explained using the analytical model described in Section \ref{mag} as follows. By considering the same Mn32 cluster illustrated in Fig. \ref{fig:CMOMAG}(c), the contribution from symmetric exchange striction can be written as,
\begin{eqnarray}
\label{eq:PES}
P_{\rm{ES}}(\theta_2) & \propto & \gamma_{41} {\bf S}_0 \cdot ({\bf S}_1+{\bf S}_2+{\bf S}_3)  \nonumber \\
       & - & \gamma_{42}{\bf S}_0 \cdot ({\bf S}_{4}+{\bf S}_{5}+{\bf S}_{6}) \nonumber \\
      & + &  \gamma_{31} {\bf S}_0 \cdot ({\bf S}_7+{\bf S}_8+{\bf S}_9) \nonumber \\
      & - & \gamma_{32}{\bf S}_0 \cdot  ({\bf S}_{10}+{\bf S}_{11}+{\bf S}_{12})  \nonumber \\
      & = & \gamma_a \cos(\frac{\pi}{3}-\theta_2)+\gamma_b\cos(\frac{\pi}{3}+\theta_2)  \nonumber \\
      & = & \frac{{1}}{2}(\gamma_a+\gamma_b)\cos(\theta_2) \nonumber \\
      & + & \frac{\sqrt{3}}{2}(\gamma_a-\gamma_b)\sin(\theta_2),
\end{eqnarray}
where $\gamma_{31},\gamma_{32},\gamma_{41},$ and $ \gamma_{42}$ are exchange striction parameters. In the linear approximation, the parameters are proportional to the first-order derivative of the corresponding $J$'s with respect to the ionic displacements.\cite{wang08} In Eq.~(\ref{eq:PES}) we have defined $\gamma_a=3(\gamma_{31}-\gamma_{42})$, and $\gamma_b=3(\gamma_{41}-\gamma_{32})$.
Likewise, the contribution from SOC can be accurately described as a DM striction effect, i.e. due to the imbalance in the DM interactions between Mn3 and the upper and lower Mn1 and Mn2 layers, respectively. This contribution can therefore be written as
\begin{eqnarray}
\label{eq:PDM} 
P_{\rm{DM}}(\theta_2) & \propto & \beta_{41} {\bf S}_0 \times ({\bf S}_1+{\bf S}_2+{\bf S}_3) \nonumber \\
      & + & \beta_{42}{\bf S}_0 \times ({\bf S}_{4}+{\bf S}_{5}+{\bf S}_{6})  \nonumber \\
      & + & \beta_{31} {\bf S}_0 \times ({\bf S}_7+{\bf S}_8+{\bf S}_9)  \nonumber \\
      & + & \beta_{32}{\bf S}_0 \times ({\bf S}_{10}+{\bf S}_{11}+{\bf S}_{12})  \nonumber \\
      & = & \beta_a \sin\left(\frac{\pi}{3}-\theta_2\right)+\beta_b\sin\left(\frac{\pi}{3}+\theta_2\right) \nonumber \\
      & = & \frac{\sqrt{3}}{2}(\beta_a+\beta_b)\cos(\theta_2) \nonumber \\
      & + & \frac{{1}}{2}(\beta_a-\beta_b)\sin(\theta_2),
\end{eqnarray}
where $\beta_{31}, \beta_{32}, \beta_{41}, $ and $ \beta_{42}$ are coupling constants 
corresponding to ${\bf D}_{31}, {\bf D}_{32}, {\bf D}_{41}, $ and $ {\bf D}_{42}$, respectively, and we have defined
$\beta_a=3(\beta_{31}+\beta_{42})$, and $\beta_b=3(\beta_{41}+\beta_{32})$. 
The contribution to the total polarization from Mn31 and Mn33 clusters could be derived in the same way.
The total polarization is then $P_{\rm{tot}}= P_{\rm{ES}}+ P_{\rm{DM}}$ , where $P_{\rm{ES}}=\sum_{i=1}^3P_{\rm{ES}}(\theta_i)$ and $P_{\rm{DM}}=\sum_{i=1}^3 P_{\rm{DM}}(\theta_i)$. 

In the average structure $\theta_1=\theta_2=\theta_3=\theta$, $\gamma_a=-\gamma_b$ and $\beta_a=\beta_b$, hence Eq. (\ref{eq:PES}) may be simplified to $P_{\rm{ES}} \propto \sin(\theta)$, and Eq. (\ref{eq:PDM}) to $P_{\rm{DM}} \propto \cos(\theta)$. The corresponding fits are also drawn in Fig. \ref{fig:polar} as lines for comparison. We can see that there is excellent agreement between the analytical expression and the results calculated {\it ab initio}. However, it is worth noting that the total polarization calculated for the optimized $\theta=4.1^\circ$ is only -0.46 mC/m$^2$, about five times \emph{smaller} than the experimental value. We will come back to this after examining the electric polarization in a modulated structure.

\begin{figure}
\centering
\includegraphics[width=3.5in]{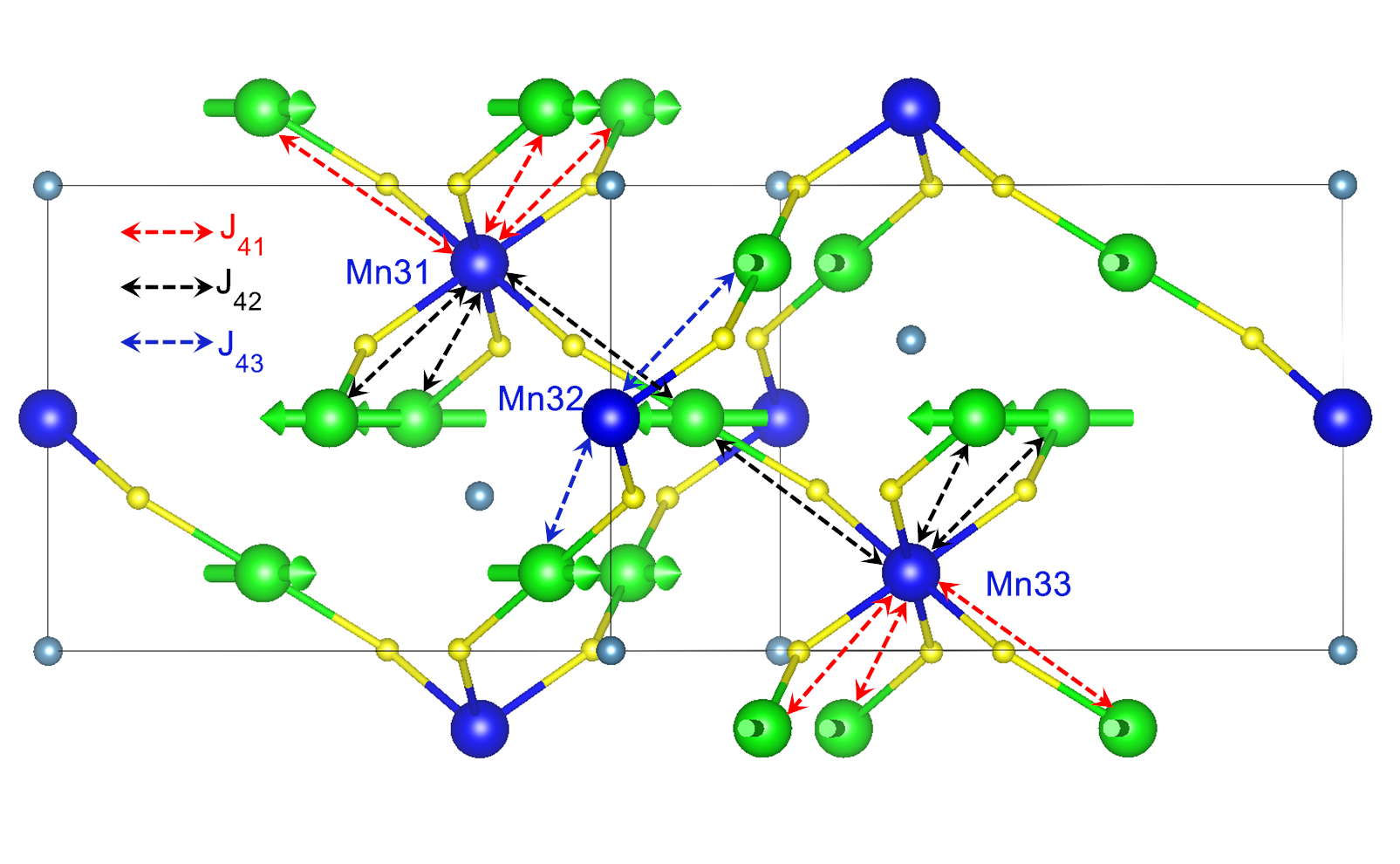}
\caption { Schematic illustrations of the splitting of the $J_4$ exchange interactions in the CM structure. The color scheme is the same as in Fig.~\ref{fig:CMOMAG}.}
\label{fig:CMO_structure}
\end{figure}

As the electric polarization is a bulk property, we cannot approximate the full modulated structure by replicating a single unit cells of interest, as we have done in the previous sections. We therefore investigate a commensurately modulated (CM) atomic structure with ${\bf q}^{\prime}_c=(0,0,2)$, constructed from the experimental data.\cite{perks12} It is worth emphasising the subtly  different approach we take here: in sections III and IV, we ``slice off'' a section of the incommensuratey modulated structure corresponding to a hexagonal conventional unit  cell, which is a supercell of the rhombohedral unit cell, and replicate it.  This results in locally exact bond lengths and angles, but also in a slight mismatch at the boundary between unit cells.  In this section, we fix the 4-dimensional fractional coordinates of the modulation to the experimental values, but we approximate the propagation vector with the commensurate value ${\bf q}^{\prime}_c=(0,0,2)$.  This also results in a hexagonal supercell of the average structure. The local atomic positions, and consequently bond lengths and angles,  are everywhere only an approximation of the real structure, but the modulation is smooth and there is no discontinuity at the unit cell boundary. The CM atomic structure is constructed based on the cell at $R_z=0$ in Fig.~\ref{fig:bondmod}. In the super-cell of this CM structure, each type of exchange interaction is split into three distinct interactions, $i.e.$ $J_i$ splits into $J_{i1}$, $J_{i2}$, $J_{i3}$. By way of example, the splitting of $J_4$ is illustrated in Fig. ~\ref{fig:CMO_structure}. 
Inversion symmetry in the CM structure is still preserved with Mn32 occupying the inversion centre, while Mn31 and Mn33 occupy two sites related by inversion symmetry.

According to the symmetry of the positions occupied by Mn3 atoms, we have $\theta_2=0$ and $\theta_1=-\theta_3$ for magnetic structures without SOC. Indeed, our calculations give values of $\theta_1=25^\circ$, $\theta_2=0$, $\theta_3=-25^\circ$. By including SOC we find $\theta_1=20^\circ$, $\theta_2=-5^\circ$, $\theta_3=-30^\circ$, where all the three $\theta$ values show a constant rotational offset of $\Delta \theta= -5^\circ$ compared to that calculated without SOC, consistent with our analysis and our calculated results in Sec. \ref{mag}. In the CM structure, a straightforward calculation according to Eq.~(\ref{eq:PES}) and Eq.~(\ref{eq:PDM}) shows
that $P_{\rm{ES}} \propto \sin(\Delta \theta)$ and $P_{\rm{DM}} \propto \cos(\Delta \theta)$. To further study this relation, the polarization calculated by fixing various $\Delta \theta$ are shown in Fig. \ref{fig:polar} (b). Here, a perfect fit to the analytical sine and cosine function can be obtained. The contributions from Mn31 and Mn33 to $P_{\rm{ES}}$ partly cancel each other due to their opposite rotation directions. The overall behaviour of the electric polarization is then very similar to that of the average structure, except for the fact that $\theta$ of the averaged structure is replaced by the global rotational offset $\Delta \theta$.  In both cases, the total polarisation results from fine tuning of $P_{\rm{ES}}$ and $P_{\rm{DM}}$.  For the {\it ab initio} optimized  $\Delta \theta=-5^\circ$, we calculate $P_{\rm{ES}}=+2.03$ mC/m$^2$ and $P_{\rm{DM}}=-2.07$ mC/m$^2$, resulting an almost exact cancellation of the  total polarization ($P_{\rm{tot}}=-0.04$ mC/m$^2$ --- two orders of magnitude smaller than the experimental value). We also performed calculations with CM atomic structures constructed from other cells of the ICM structure. Polarizations from different CM structures show similar modulation with $R_z$ as in Fig.~\ref{fig:bondmod}, while the total polarization calculated by averaging the polarizations from each CM structure is consistent with that obtained using the CM cell from $R_z=0$. 

One probable reason for this coincidental cancellation is that in our calculation  the atomic structure is only approximated, as explained above, and in not fully relaxed. We therefore repeated our \emph{ab-intio} calculations with relaxed atomic positions of the CM structure but keeping the experimental lattice parameters. After relaxation, $\Delta \theta$ slightly increases to $-5.6^\circ$ and the total polarization is greatly enhanced to $P_{\rm{tot}}=-1.19$ mC/m$^2$, much closer to the experimental value.  It is worth reminding that the \emph{sign} agreement between measured and calculated polarizations is extremely difficult to determine, since it requires establishing experimentally the absolute \emph{chirality} of the magnetic structure.  The atomic relaxation changes the Mn-Mn bonds and hence the corresponding coupling parameters $\gamma$ and $\beta$, leading to an increase in the contribution from the DM interaction relative to that from exchange striction. Another possibility is that the calculated value of  $\Delta \theta$ is too small, possibly due to an underestimation of the DM interaction. If we enforce $\Delta \theta = -10^\circ$, the contribution from exchange striction  prevails and a total polarization of $P_{\rm{tot}}= 2.4$ mC/m$^2$ is obtained, which is in good agreement with experiments \cite{johnson12}. From this analysis, it is clear that the DM interaction plays a key role in producing macroscopic electric polarization. In the absence of DM interaction, $\Delta \theta=0$, and both $P_{\rm{ES}}$ and $P_{\rm{tot}}$ vanish. When the DM interaction is present, a net electric polarization arises with contributions from both $P_{\rm{ES}}$ and $P_{\rm{DM}}$.
This conclusion, based on the analysis of our calculated data, is robust and largely insensitive to the calculation details.

%

\section{Conclusions}
We have presented a series of \textit{ab initio} calculations on multiferroic CaMn$_7$O$_{12}$, based on the experimentally determined modulated crystal structure and designed to provide insight into the electronic and magnetic structures and the nature of the magnetically induced electrical polarization of this material.  We also presented a series of analytical models expressed in terms of a small number of parameters, which provide excellent fits to the  \emph{ab-intio} calculations.
By investigating the $3d$ orbital occupation using a PDOS analysis, we confirmed the $3d$ $e_g$ orbital modulation of Mn2 atom previously proposed. We further investigated the detailed magnetic structure
of the Mn3 (Mn$^{4+}$) sublattice in the multiferroic AFM1 phase and its interactions with the Mn$^{3+}$-containing Mn1 and Mn2 sublattices. We find that  the Mn3 magnetic structure  is  anharmonically modulated  due to orbitally-mediated coupling with the structural modulation --- a key ingredient in stabilising the observed magneto-orbital helices. Finally, we studied the magnetoelectric polarization  both in the average (unmodulated) structure and in a commensurate approximation of the modulated structure. The Dzyaloshinskii-Moriya interaction was found to play a key role in producing magnetoelectric polarization, while the total polarisation results from a delicate interplay of inverse DM and exchange-striction contributions.  

\begin{acknowledgments}
This work was funded by an EPSRC grant, number EP/J003557/1, entitled ``New Concepts
in Multiferroics and Magnetoelectrics". Calculations were performed at the Oxford
Supercomputing Centre and at the Oxford Materials Modelling Laboratory.
\end{acknowledgments}




\end{document}